\documentclass[aps,twocolumn,superscriptaddress,floatfix,reprint]{revtex4-1}
\usepackage{graphicx}
\usepackage{epstopdf}

\usepackage{amsmath}
\usepackage{bm}
\usepackage{amssymb}
\usepackage{xcolor}
\renewcommand{\vec}[1]{\textbf{#1}}

\newcommand{\col}[2]{\left (\arraycolsep=1.4pt
\begin{array}{c}
#1\\
#2\\
\end{array}
\right)\arraycolsep=1.4pt}
\newcommand{\F}{{\rm F}}
\renewcommand{\S}{{\rm S}}
\newcommand{\N}{{\rm N}}
\newcounter{fignum}
\begin{document}

\title{Interference phenomena in Josephson junctions with ferromagnetic bilayers:
Spin-triplet correlations and resonances}

\author{Danilo Nikoli\' c}%
\affiliation{Department of Physics, University of Belgrade, Studentski trg 12, 11158 Belgrade, Serbia}
\affiliation{Fachbereich Physik, Universit\" at Konstanz, D-78457 Konstanz, Germany}

\author{Mihajlo Vanevi\' c}%
\affiliation{Department of Physics, University of Belgrade, Studentski trg 12, 11158 Belgrade, Serbia}

\author{Alexander I. Buzdin}%
\affiliation{University Bordeaux, LOMA UMR-CNRS 5798, F-33405 Talence Cedex, France}
\affiliation{World-Class Research Center “Digital Biodesign and Personalized Healthcare”, Sechenov First Moscow State Medical University, Moscow 119991, Russia}

\author{Zoran Radovi\' c}%
\affiliation{Department of Physics, University of
Belgrade, Studentski trg 12, 11158 Belgrade, Serbia}
\affiliation{Serbian Academy of Sciences and Arts, Kneza Mihaila 35, 11000, Belgrade, Serbia}

\begin{abstract}
We study the Josephson effect in planar $\S\F_1\F_2\S$ junctions that consist of conventional $s$-wave superconductors (S) connected by two metallic monodomain ferromagnets ($\F_1$ and $\F_2$) with arbitrary transparency of interfaces. We solve the scattering problem in the clean limit based on the Bogoliubov - de Gennes equation for both spin-singlet and odd in frequency spin-triplet pairing correlations. We calculate numerically the Josephson current-phase relation $I(\phi)$. While the first harmonic of $I(\phi)$ is completely generated by spin-singlet and short-range spin-triplet superconducting correlations, for noncollinear magnetizations of ferromagnetic layers the second harmonic has an additional long-range spin-triplet component. Therefore, for a strong ferromagnetic influence, the long-range spin-triplet contribution to the second harmonic dominates. We find an exception due to the geometric resonance for equal ferromagnetic layers when the first harmonic is strongly enhanced. Both first and second harmonic amplitudes oscillate with ferromagnetic layer thicknesses due to $0-\pi$ transitions. We study the influence of interface transparencies and find additional resonances for finite transparency of the interface between ferromagnetic layers. 
\end{abstract}



\date{\today}

\maketitle

\section{Introduction}
\label{sec:Intro}
The interplay between superconductivity and magnetism in proximity heterostructures~\cite{Bulaevskii1977,Buzdin1984,Radovic1988,Radovic1991} has been attracting considerable interest for decades, see for example Refs.~\cite{Tedrow1994,Jiang_1995,Obi_1999,Ryazanov_2001,Kontos_2001,Obi_2005,buzdin_proximity_2005, bergeret_odd_2005,  lyuksyutov_ferromagnetsuperconductor_2005, golubov_current-phase_2004,linder_superconducting_2015,eschrig_2015}. 
Remarkably, odd in frequency spin-triplet pairing correlations may occur in $\S\F\S$ Josephson structures comprised of superconductors with spin-singlet pairing and a metallic ferromagnet~\cite{bergeret_long-range_2001,Kadigrobov2001}. In the case of a homogeneous ferromagnet, the triplet pair amplitude has zero total spin projection on the magnetization axis. This amplitude, as well as  the spin-singlet amplitude, decay over the short length scale determined by the exchange energy $h$ in the ferromagnet. The characteristic coherence length in ferromagnet is given by $\xi_F=\hbar v_F/h$ and $\xi_F=\sqrt{\hbar D/h}$ ($D=v_F\ell/3$ is the diffusion coefficient with $\ell$ being the electronic mean free path) in the clean and diffusive limit, respectively. The situation is quite different for an inhomogeneous ferromagnet where spin-triplet pair amplitudes with $\pm 1$ total spin projection on the magnetization axis emerge~\cite{bergeret_long-range_2001}. These amplitudes decay on substantially larger length scales determined by temperature, $\xi_F=\hbar v_F/(k_B T)$ in the clean and $\xi_F=\sqrt{\hbar D/(k_B T)}$ in the diffusive limit~\cite{bergeret_odd_2005}.

A simple realization of a Josephson junction with an inhomogeneous ferromagnet is the $\S\F_1\F_2\S$ heterostructure with two monodomain ferromagnets having noncollinear in-plane magnetizations~\cite{blanter_supercurrent_2004,pajovic_josephson_2006,crouzy_josephson_2007,trifunovic_josephson_2011,trifunovic_long-range_2011,melnikov_interference_2012,richard_superharmonic_2013,knezevic_signature_2012,kawabate_2013, hikino_long-range_2013, meng_long-range_2016,meng2019}. However, in such proximity structures the long-range spin-triplet component of the supercurrent consists only of even harmonic amplitudes~\cite{trifunovic_josephson_2011,trifunovic_long-range_2011,melnikov_interference_2012,richard_superharmonic_2013}. In the case of strong ferromagnets the short-range components are suppressed and the second harmonic is dominant in the current-phase relation. Odd harmonic amplitudes can be long-ranged only in heterojunctions with three or more ferromagnetic layers~\cite{volkov_odd_2003,lofwander_interplay_2005,braude_fully_2007,houzet_long_2007,fominov_josephson_2007,sperstad_josephson_2008,konschelle_nonsinusoidal_2008,trifunovic_long-range_2010, volkov_odd_2010,alidoust_spin-triplet_2010,Samokhvalov2014, halterman_charge_2015,Chen2021}. 

Note that the anharmonic current-phase relation can be expanded as $I(\phi) = I_1 \sin \phi + I_2\sin 2\phi +\dots$, where the $n$th harmonic amplitude $I_n$ corresponds to the phase-coherent transport of $n$ Cooper pairs~\cite{trifunovic_long-range_2011}. Junctions with a pure second harmonic exhibit degenerated ground states for $\phi=0$ and $\pi$ at the so-called $0-\pi$ transition~\cite{Radovic1991,Stoutimore2018}. A small contribution of  other harmonics lifts degeneracy and leads to the coexistence of stable and metastable $0$ and $\pi$ states~\cite{radovic_coexistence_2001,radovic_josephson_2003,Buzdin_dirty2005,Barsic2006}.

A long-ranged supercurrent has been observed in  Nb Josephson junctions with Ni- and Co-based ferromagnetic multilayers~\cite{khaire_observation_2010,Lahabi2017,Glick2018,Kapran2020, Aguilar2020}. An enhanced second harmonic in the long-ranged supercurrent has been observed in mesa-heterostructures of cuprate superconductors and ferromagnetic bilayers of  manganite and ruthenate~\cite{ovsyannikov_triplet_2013}, while a pure second harmonic has been observed in $\rm NbN/GdN/NbN$ junctions~\cite{pal_pure_2014}.

A dominant second harmonic, $I_2\gg I_1$, can be realized in the regime of highly asymmetric $\S\F_1\F_2\S$ junctions~\cite{trifunovic_josephson_2011,trifunovic_long-range_2011, melnikov_interference_2012,richard_superharmonic_2013}. The physical picture behind this effect is the following: At the SF interface the exchange field generates spin-triplet correlations with $0$ spin projection. Penetrating into the next ferromagnetic layer with misoriented magnetization they mix forming long-range spin-triplet correlations with $\pm 1$ spin projection. Therefore, for a fully developed spin-triplet proximity effect one of two ferromagnetic layers should be sufficiently thin to provide a large short-range spin-triplet amplitude with zero spin projection at the interface between ferromagnetic layers in order to generate large long-range spin-triplet amplitudes with $\pm 1$ spin projection. The other ferromagnetic layer should be sufficiently thick to filter out the short-range correlations~\cite{Volkov2020}.

In this paper we study the Josephson effect in clean planar (three-dimensional) $\S\F_1\F_2\S$ junctions that consist of conventional $s$-wave superconductors and two metallic monodomain ferromagnets (equal strength and different thicknesses) with arbitrary transparency of the interfaces. We calculate numerically the Josephson current-phase relation $I(\phi)$ by using the Bogoliubov-de Gennes formalism. In particular, we calculate the first and second harmonic amplitudes, $I_1$ and $I_2$. The long-range second harmonic is well-pronounced for an overall strong ferromagnetic influence due to the contribution of the odd-frequency spin-triplet correlations with $\pm 1$ spin projection. However, for equal thicknesses of ferromagnetic layers the spin-singlet contribution to the first harmonic is dominant due to geometric resonances even for a strong ferromagnetic influence. In a previous paper the results for linear (one-dimensional) $\S\F_1\F_2\S$ structures were illustrated only for equal ferromagnetic layers and the influence of the long-range spin-triplet correlations was completely hidden~\cite{pajovic_josephson_2006}. In subsequent papers using the same approach~\cite{meng_long-range_2016,meng2019}, the interplay between the geometric resonances and spin-triplet correlations was not studied explicitly. Here, we focus on this subject.

Both the first and the second harmonic oscillate with ferromagnetic layer thicknesses due to the $0-\pi$ transitions. For finite transparency of interfaces the supercurrent is suppressed, with higher harmonics being more affected. A lower transparency of the interface between ferromagnetic layers, where the long-range spin-triplet correlations are generated, has a nontrivial impact on the interference phenomena: For certain thicknesses of the ferromagnetic layers we find additional geometric resonances. 

The paper is organized as follows. In Sec.~\ref{sec:Model} we present the model and the solution. In Sec.~\ref{sec:Results} we present and discuss the numerical results for the Josephson current and harmonic amplitudes. Finally, the concluding remarks are given in Sec.~\ref{sec:Conclusion}.

\section{Model}
\label{sec:Model}
\subsection{The Bogoliubov-de Gennes equations for $\S\F_1\F_2\S$ heterojunctions}
We consider a clean planar (three-dimensional) $\S\mathcal{I}_1\F_1\mathcal{I}_2\F_2\mathcal{I}_3\S$ heterojunction that consists of two superconductors (S), two uniform monodomain ferromagnetic layers ($\F_1$ and $\F_2$), and three nonmagnetic interfacial potential barriers between metallic layers $(\mathcal{I}_1- \mathcal{I}_3)$, depicted in Fig.~\ref{fig:Schema}.  Superconductors are described in the framework of BCS formalism, while for ferromagnets we use the Stoner model with a spatially-dependent energy shift $2h(\vec{r})$ between the spin subbands. The model and methods are the same as in the previous papers~\cite{pajovic_josephson_2006,meng_long-range_2016, meng2019}.

Electronlike and holelike quasiparticles with energy $E$ and spin projection $\sigma = \uparrow,\downarrow$
are described by $u_\sigma(\vec{r})$ and $v_\sigma(\vec{r})$, respectively, where $\vec{r}$ is the spatial coordinate. Using the four-component wave function $\Psi(\vec{r}) = [ u_\uparrow(\vec{r}), u_\downarrow(\vec{r}), v_\uparrow(\vec{r}), v_\downarrow(\vec{r}) ]^T$, the Bogoliubov-de Gennes equation has the following form: 
\begin{figure}[t!]
	\includegraphics[width=\columnwidth]{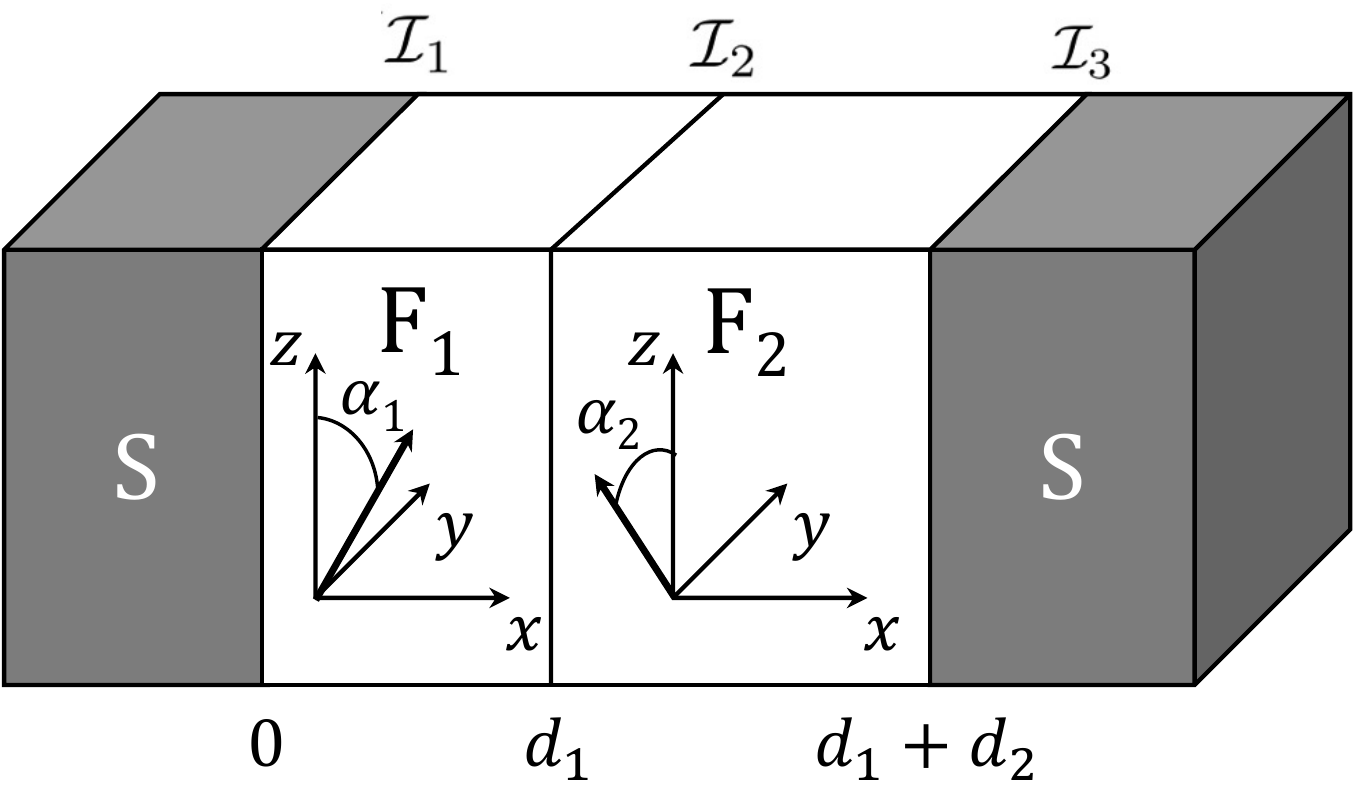}
	\caption{
		{Schematic representation of an $\S \F_1 \F_2 \S$ junction. Two ferromagnetic layers $\F_1$ and $\F_2$ of thicknesses $d_1$ and $d_2$, respectively, are coupled to two superconducting electrodes (S).
			The magnetization vectors lie in the $yz$ plane at angles $\alpha_1$ and $\alpha_2$ with respect to the $z$ axis. The insulating interfaces between the superconducting and ferromagnetic layers are denoted as $\mathcal{I}_1-\mathcal{I}_3$}.}
	\label{fig:Schema}
\end{figure}
\begin{equation}
\check {\mathcal{H}}\Psi(\vec{r}) = E\Psi(\vec{r})
\label{eqn:BdG},
\end{equation}
with $\check{\mathcal{H}}$ being a $2\times 2$ matrix in particle-hole space
\begin{equation}
\check {\mathcal{H}}=
{
\left (\arraycolsep=1.6pt
\begin{array}{cc}
\hat{H}(\vec{r}) & \hat{\Delta} \\
 \hat{\Delta}^* &  -\hat{H}(\vec{r}) \\
\end{array}
\right)
        },
\end{equation}
where each block itself is a $2\times 2$ matrix in spin space, such that, $\hat{H}(\vec{r}) = H_0(\vec{r})\hat{1} -
 h(\vec{r})\sin[\alpha(\vec{r})]\hat{\tau}_2 - h(\vec{r})\cos[\alpha(\vec{r})]\hat{\tau}_3$
 and $\hat{\Delta}(\vec{r}) = \Delta (\vec{r}) \hat{\tau}_1$. Here, $\hat{\tau}_i$ are
 Pauli matrices, $\hat{1}$ is the unity matrix, and $H_0(\vec{r}) = -\hbar^2\nabla^2/2m +
 W(\vec{r}) + U(\vec{r}) - \mu$. The chemical potential is denoted by $\mu$, $W(\vec{r}) =
 \sum_i W_i\delta(x -x_i)$ is the potential of the barriers at the interfaces, and $U(\vec{r})$
 is the electrostatic potential. The $x$ axis is chosen to be perpendicular to the layers,
 whereas $x_1 = 0$, $x_2=d_1$, and $x_3 = d_1 +d _2$ are coordinates of the interfaces.
 At zero temperature the difference $\mu - U(\vec{r})$ is equal to the Fermi energy
 $E_F$. The in-plane $(y - z)$ magnetizations of the two F layers are not
 collinear in general, and the magnetization orientation is defined by the angle
 $\alpha(\vec{r})$ with respect to the $z$ axis. We choose $\alpha(\vec{r}) = \alpha_1$
 for $0 < x < d_1$ in $\F_1$, and $\alpha(\vec{r}) = \alpha_2$ for $d_1 < x < d_1 + d_2$ in $\F_2$ (see Fig.~\ref{fig:Schema}).

For simplicity, we assume equal magnitudes of the exchange field in the ferromagnets,
$h(\vec{r}) = h \Theta (x)\Theta (d_1 + d_2 - x)$ where $\Theta(x)$ stands for the Heaviside
step function. We also assume that the effective electron mass $m$ is constant throughout the layers and that the mean Fermi energy of the ferromagnets, $E_F = (E_F^\uparrow + E_F^\downarrow)/2$, is equal to the Fermi energy of superconductors. However, the influence of the mismatch of effective electron masses and Fermi energies is similar to the influence of finite interfacial transparency: an increase of the normal refection~\cite{radovic_josephson_2003,Zutic2000,KuprianovLukichev}.

We take the pair potential $\Delta(\vec{r})$ in the form
\begin{equation}\label{eqn:pair_potential}
\Delta(\vec{r}) = \Delta [e^{i\phi_1}\Theta (-x) + e^{i\phi_2}\Theta(x - d_1 - d_2)],
\end{equation}
where $\Delta$ is the bulk superconducting gap. The macroscopic phase difference across the
junction is $\phi =\phi_1 -\phi_2$. The temperature dependence of $\Delta$ is given by
$\Delta (T) = \Delta (0)\tanh (1.74\sqrt{T_c/T - 1})$ with $\Delta(0)$ being the pair potential at zero temperature and $T_c$ the critical temperature~\cite{muhlschlegel_1959}. In general, $\Delta(\vec{r})$ should be determined self-consistently~\cite{halterman_charge_2015,meng_long-range_2016}. However, for simplicity we use the stepwise pair potential given in Eq.~(\ref{eqn:pair_potential}), since
self-consistency will not alter our results qualitatively~\cite{knezevic_signature_2012}. 

\subsection{Solution of the scattering problem}
The parallel component of the wave vector, $\vec{k}_\parallel$, is conserved due to the
translational invariance of the junction in the direction perpendicular to the $x$ axis.
Consequently, the wave function can be written in the form
\begin{equation}
\Psi(\vec{r}) = \psi(x)e^{i\vec{k}_\parallel \cdot \vec{r}},
\end{equation}
where $\psi(x) =  [ u_\uparrow(x), u_\downarrow(x), v_\uparrow(x), v_\downarrow(x) ]^T$
satisfies the following boundary conditions:
\begin{equation}
\psi(x)|_{x_i + 0} = \psi(x)|_{x_i - 0} = \psi(x_i),
\label{eqn:BC1}
\end{equation}
\begin{equation}
\frac{d\psi(x)}{dx} \bigg |_{x_i+0} - \frac{d\psi(x)}{dx} \bigg |_{x_i-0} = Z_ik_F\psi(x).
\label{eqn:BC2}
\end{equation}
Here, $Z_i = 2mW_i/\hbar^2 k_F$ $(i = 1,2,3)$ are parameters~\cite{BTK-82}
that measure the transparency of the interfaces (i.e., barrier strengths) located
at $x_i = 0, d_1, d_1 + d_2$, and $k_F = \sqrt{2mE_F/\hbar^2}$ is the Fermi wave vector.


The four independent solutions of the scattering problem for Eq.~(\ref{eqn:BdG}) correspond
to the four types of quasiparticle injection processes: an electronlike or a holelike
quasiparticle injected from either the left or the right superconducting electrode.
When an electronlike quasiparticle is injected from the left, the solutions of Eq.~(\ref{eqn:BdG}) for the left superconductor ($x < 0$) written in the compact form are 

\begin{align}
\col{u_\sigma}{v_{\bar{\sigma}}} = &
\col{\bar{u} e^{i\phi_1/2}}{\bar{v}e^{-i\phi_1/2}} e^{ik^+x}
+ b_{1\sigma} \col{ \bar{u} e^{i\phi_1/2}}{\bar{v}e^{-i\phi_1/2}} e^{-ik^+x}
\notag
\\
 +& a_{1\sigma} \col{\bar{v}e^{i\phi_1/2}}{\bar{u} e^{-i\phi_1/2}} e^{ik^-x},
\label{IN1}
\end{align}
and for the right superconductor $(x>d_1 + d_2)$
\begin{equation}
\col{u_\sigma}{v_{\bar{\sigma}}} =
c_{1\sigma} \col{\bar{u}e^{i\phi_2/2}}{\bar{v}e^{-i\phi_2/2}} e^{ik^+x}
+ d_{1\sigma}\col{\bar{v}e^{i\phi_2/2}}{\bar{u}e^{-i\phi_2/2}} e^{-ik^-x},
\label{OUT1}
\end{equation}
where $\bar{\sigma}$ is opposite to $\sigma=\uparrow\downarrow$,
$\bar u =\sqrt{(1+\Omega/E)/2}$, $\bar v = \sqrt{(1-\Omega/E)/2}$, and $\Omega = \sqrt{E^2 - \Delta^2}$.
Constants  $a_{1\sigma}, b_{1\sigma}, c_{1\sigma}$, and $d_{1\sigma}$ correspond to Andreev and normal
reflections, direct transmission, and nonlocal Andreev reflection, respectively. For the left
ferromagnetic layer $\F_1~(0<x<d_1)$ solutions of Eq.~(\ref{eqn:BdG}) are
\begin{align}
\col{u_\uparrow}{u_\downarrow} &=
c_1 \col{i\cos{\frac{\alpha_1}{2}}}{-\sin{\frac{\alpha_1}{2}}} e^{iq^+_\uparrow x}
+ c_2 \col{i\cos{\frac{\alpha_1}{2}}}{-\sin{\frac{\alpha_1}{2}}} e^{-iq^+_\uparrow x}
\notag
\\
&
+ c_3 \col{i\sin{\frac{\alpha_1}{2}}}{\cos{\frac{\alpha_1}{2}}} e^{iq^+_\downarrow x}
+ c_4\col{i\sin{\frac{\alpha_1}{2}}}{\cos{\frac{\alpha_1}{2}}} e^{-iq^+_\downarrow x},
\label{FF1}
\end{align}
\begin{align}
\col{v_\uparrow}{v_\downarrow} &=
c_5 \col{i\cos{\frac{\alpha_1}{2}}}{-\sin{\frac{\alpha_1}{2}}} e^{iq^-_\uparrow x}
+ c_6 \col{i\cos{\frac{\alpha_1}{2}}}{-\sin{\frac{\alpha_1}{2}}} e^{-iq^-_\uparrow x}
\notag
\\
& + c_7 \col{i\sin{\frac{\alpha_1}{2}}}{\cos{\frac{\alpha_1}{2}}} e^{iq^-_\downarrow x}
+ c_8 \col{i\sin{\frac{\alpha_1}{2}}}{\cos{\frac{\alpha_1}{2}}} e^{-iq^-_\downarrow x}.
\label{FL}
\end{align}
Solutions for the right ferromagnetic layer $\F_2~(d_1<x<d_1+d_2)$ can be obtained by substituting $\alpha_1\rightarrow\alpha_2$, with a new set of constants $c'_1,\ldots, c'_8$.

When a holelike quasiparticle is injected from the left, the solutions of Eq.~(\ref{eqn:BdG}) for the left
superconductor $(x<0)$ are given by
\begin{align}
\col{u_\sigma}{v_{\bar{\sigma}}} =&
\col{\bar{v}e^{i\phi_1/2}}{\bar{u} e^{-i\phi_1/2}} e^{-ik^-x}
+ b_{2\sigma} \col{\bar{v}e^{i\phi_1/2}}{\bar{u} e^{-i\phi_1/2}} e^{ik^-x}
\notag
\\
 + &
a_{2\sigma} \col{ \bar{u} e^{i\phi_1/2}}{\bar{v}e^{-i\phi_1/2}} e^{-ik^+x},
\label{IN2}
\end{align}
while for the right superconductor $(x>d_1+d_2)$
\begin{equation}
\col{u_\sigma}{v_{\bar{\sigma}}}
=
c_{2\sigma} \col{\bar{v}e^{i\phi_2/2}}{\bar{u}e^{-i\phi_2/2}} e^{-ik^-x}
+ d_{2\sigma}\col{\bar{u}e^{i\phi_2/2}}{\bar{v}e^{-i\phi_2/2}} e^{ik^+x}.
\label{OUT2}
\end{equation}
Constants $a_{2\sigma}, b_{2\sigma}, c_{2\sigma}$, and  $d_{2\sigma}$ describe analogous processes as in
the case of an injected electronlike quasiparticle given earlier.

Solutions for ferromagnetic layers in the case a holelike quasiparticle injected from the left can be obtained by substituting $c_i \rightarrow C_i$ and $c'_i\rightarrow C'_i$ in solutions for the case of an injected electronlike quasiparticle.
The longitudinal $x$ - components of the wave vectors in the superconductors are given by
\begin{equation}
\label{eqn:k_vector}
k^{\pm}=\sqrt{\frac{2m}{\hbar^2}(E_F\pm\Omega) - \vec{k}_\parallel^2},
\end{equation}
while their counterparts in the ferromagnetic layers read
\begin{equation}
q^{\pm}_{\sigma} = \sqrt{\frac{2m}{\hbar^2}(E_F\pm E+\rho_\sigma h) - \vec{k}_\parallel^2}.
\end{equation}
The sign $\pm$ in the superscript corresponds to the sign of the quasiparticle energy, whereas $\rho_\sigma=+1~(-1)$ is related to the spin projection $\sigma = \uparrow(\downarrow)$.

Solutions for quasiparticles with opposite spin orientations  are nontrivially coupled: in the superconductors, Eqs.~\eqref{IN1} and \eqref{OUT1} and Eqs.~\eqref{IN2} and \eqref{OUT2}, as well as in the ferromagnets,  Eqs.~\eqref{FF1} and \eqref{FL}. In that manner, both the usual and spin-flip Andreev reflections are taken into account.
	
All the unknown 48 coefficients in the above solutions, in both electronlike and holelike
scattering problems, are determined from the boundary conditions, Eqs.~(\ref{eqn:BC1}) and (\ref{eqn:BC2}), at the three interfaces.

\subsection{The Josephson current}
The Josephson current can be calculated from the linear superposition of amplitudes the normal and anomalous Andreev reflections~\cite{Furusaki1991}, $a_{1\sigma}$ and $a_{2\sigma}$, 
\begin{equation}
I(\phi) = \frac{e\Delta}{2\hbar}\sum_{\sigma,\vec{k}_\parallel, \omega_n}
\frac{k_BT}{2\Omega_n}(k^+_n+k^-_n)
\left[
\frac{a_{1\sigma n}(\phi)}{k^+_n} - \frac{a_{2\sigma n}(\phi)}{k^-_n}
\right ].
\label{eqn:JC}
\end{equation}
Here, $\phi = \phi_1-\phi_2$ is the superconducting phase difference, and $a_{1\sigma n}, a_{2\sigma n}$, $k^\pm_n$, and $\Omega_n=\sqrt{\omega_n^2+\Delta^2}$ are obtained from the corresponding quantities shown in the previous section by performing the analytic continuation, $E\to i\omega_n$. The Matsubara frequencies are $\omega_n=(2n+1)\pi k_BT$, with $n=0, \pm 1, \pm 2, ...$ and the temperature $T$.

For nonmagnetic (SNS and SIS) Josephson junctions $a_{1\sigma}$ and $a_{2\sigma}$ are $\sigma$ independent and related by particle-hole symmetry, $a_1(\phi)=a_2(-\phi)$. However, for SFS junctions (with homogeneous/inhomogeneous magnetization), when the odd-frequency spin-triplet correlations (short/long range) are generated, the amplitudes $a_{1\sigma}$ and $a_{2\sigma}$ are $\sigma$ dependent. In that way, the spin-mixing processes are included.

Performing a summation over $\vec{k}_\parallel$ by employing $\sum_{\vec{k}_\parallel}\rightarrow \mathcal{A}(2\pi)^{-2}\int d^2k$, we obtain
\begin{equation}
\begin{split}
I (\phi)=\  &\frac{\Delta\pi}{R_Ne}k_BT\int_0^{\pi/2}d\theta \sin \theta \cos \theta \times
\\
&\times \sum_{\sigma\omega_n}\frac{k^+_n+ k^-_n}{4\Omega_n}\bigg(\frac{a_{1\sigma n}(\phi)}{k^+_n} -\frac{a_{2\sigma n}(\phi)}{k^-_n}\bigg).
\end{split}
\label{eqn:JC_3D}
\end{equation}
Here, $R_N=2\pi^2\hbar/\mathcal{A}e^2k_F^2$ with $\mathcal{A}$ being the cross section of the junction and
we assume $k_\parallel=k_F\sin\theta$, since we deal with standard BCS superconductors, where $\Delta/E_F \sim 10^{-3}- 10^{-4}$.

In general, the current-phase relation is an anharmonic $2\pi$-periodic function and can be expanded as 
\begin{equation}
\label{eqn:expn}
I(\phi) = I_1 \sin \phi + I_2\sin 2\phi +\dots,
\end{equation}
 where the $n$th harmonic amplitude $I_n$ corresponds to the phase-coherent transport of $n$ Cooper pairs. 

\section{Results and discussion}
\label{sec:Results}
We illustrate our results on $\S\F_1\F_2\S$ planar junctions with relatively weak ferromagnets
$h/E_F = 0.1$ at low temperature $T/T_c = 0.1$.
The ferromagnetic coherence length is
$\xi_F = \hbar v_F/h = 20\, k_F^{-1}$. Superconductors are characterized by the bulk pair
potential at zero temperature $\Delta(0)/E_F = 10^{-3}$ which corresponds to the
superconducting coherence length $\xi_S(0) = \hbar v_F/\pi\Delta(0) = 636\, k_F^{-1}$.
The Josephson current is normalized to $\pi\Delta/eR_N$ as usual [see Eq.~(\ref{eqn:JC_3D})]. The total thickness of the ferromagnetic bilayer is kept constant, $d_1+d_2=1000k_F^{-1}=50\xi_F=1.57\xi_S(0).$
\subsection{Fully transparent interfaces}
\begin{figure}[t!]
	\centering
	\includegraphics[scale=1]{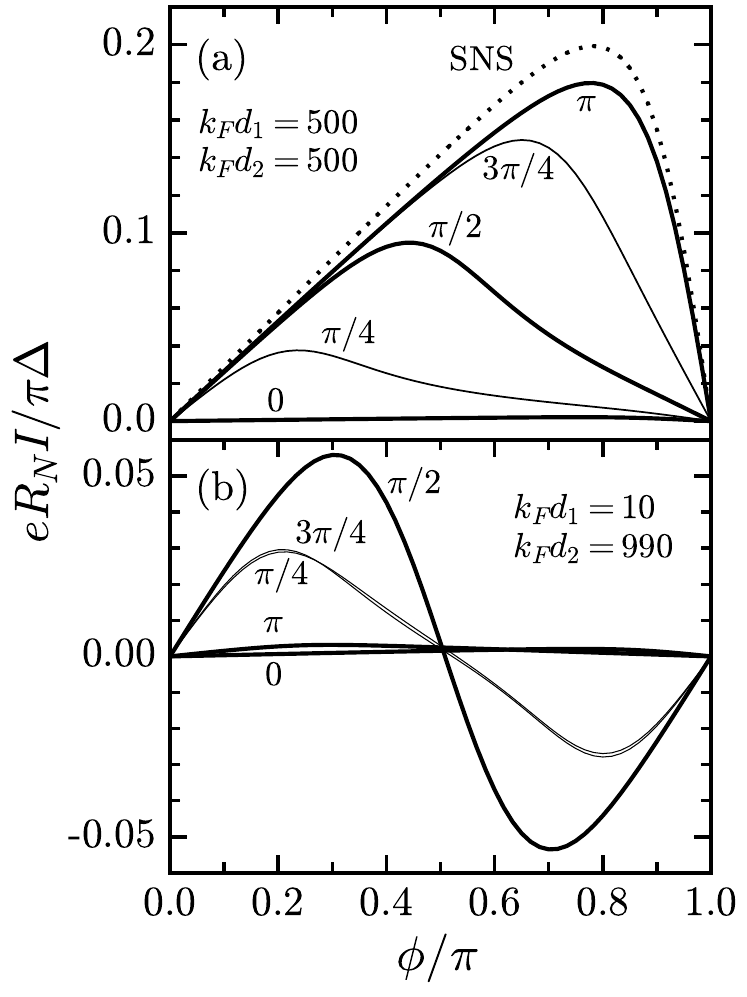}
	\caption{The Josephson current in $\S\F_1\F_2\S$ junctions as a function of the
		superconducting phase difference $\phi$ for the ferromagnetic layer
		thicknesses (a) $d_1 = d_2 = 500 k_F^{-1}$ and (b) $d_1 = 10
		k_F^{-1}$, $d_2 = 990 k_F^{-1}$, for $h/E_F = 0.1$, $T/T_c = 0.1$,
		and different relative angles between the magnetizations:
		$\alpha_r=0$, $\pi/4$, $\pi/2$, $3\pi/4$, $\pi$.
		The Josephson current for SNS junction ($h=0$)
		with the thickness $d_1 + d_2 = 1000 k_F^{-1}$ is shown in the
		panel (a) for comparison (dotted line).}
	\label{fig:IvsPhi}
\end{figure}
\begin{figure}[!]
	\begin{center}
		\includegraphics[scale=0.97]{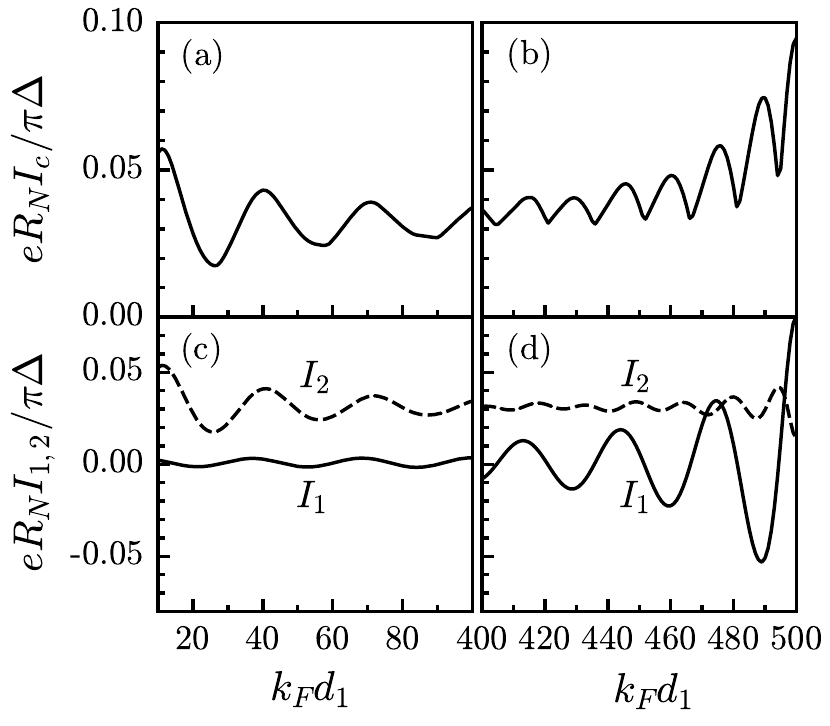}
	\end{center}
	\caption{
		The critical current in $\S\F_1\F_2\S$ junctions with mutually orthogonal magnetizations $\alpha_r=\pi/2$ and fully transparent interfaces, $Z_1=Z_2=Z_3=0$, shown as a function of the
		$\F_1$ layer thickness $d_1$: (a) thin and
		(b) thick $\F_1$ layer. The total thickness is
		$d_1+d_2 = 1000k_F^{-1}$.
		The amplitudes of the first (solid line) and the second
		harmonic (dashed line) of the Josephson current are shown
		in (c) and (d).
	}
	\label{fig:IcHarmonics1}
\end{figure}

\vfill
The current-phase relation  in a junction with fully transparent interfaces, $Z_1=Z_2=Z_3=0$, for various values of the relative angle between magnetizations, $\alpha_r=\alpha_1-\alpha_2$, and equal thicknesses of the ferromagnetic layers is shown in Fig.~\ref{fig:IvsPhi}(a). It can be seen that the current is completely suppressed for the parallel magnetizations, $\alpha_r=0$, and increases almost monotonously with a misorientation of magnetizations up to $I(\phi)$ of the corresponding $\S\N\S$ junction $(h=0)$ for the antiparallel magnetizations, $\alpha_r=\pi$, which has been observed experimentally~\cite{robinson_enhanced_2010}. The current-phase relation is a practically universal function of the ferromagnetic influence, which is measured by the product of thickness and the exchange field strength, $d\cdot h$~\cite{radovic_josephson_2003}. This explains the cancellation of ferromagnetic influence in the case of  equal thicknesses and equal strengths of the ferromagnets. However, in that case no significant influence of the triplet correlations  was found even for noncollinear magnetizations~\cite{pajovic_josephson_2006}. This we explain now by a dominant first harmonic due to the geometric resonance effect [see Fig.~\ref{fig:IcHarmonics1}(d)].

A dominant second harmonic can be seen in Fig.~\ref{fig:IvsPhi}(b) for highly unequal thicknesses of the ferromagnetic layers, $k_Fd_1=10$ and $k_Fd_2=990$, and noncollinear magnetizations. In contrast to the case of equal ferromagnetic layers, the critical current is not a monotonous function of the misorientation angle $\alpha_r$. It almost vanishes for $\alpha_r=0,\pi$ and reaches the maximum for $\alpha_r=\pi/2$. This is a manifestation of the long-range spin-triplet proximity effect in ferromagnetic bilayers where the first harmonic is suppressed and the phase-coherent transport of two Cooper pairs becomes dominant~\cite{trifunovic_josephson_2011, trifunovic_long-range_2011, melnikov_interference_2012, knezevic_signature_2012, richard_superharmonic_2013}.

\vfill
To illustrate the role of ferromagnetic bilayer asymmetry, we calculate the critical current $I_c$ and the amplitudes of the first and the second harmonic, $I_1$ and $I_2$, as functions of the $\F_1$ layer thickness, $d_1$, keeping the total thickness constant, $k_F(d_1+d_2)=1000$. The relative angle between the magnetizations is $\alpha_r=\pi/2$ (the strongest effect of spin-triplet correlations) and the interfaces are fully transparent, $Z_1 = Z_2 = Z_3 = 0$.  Results are shown in Fig.~\ref{fig:IcHarmonics1}. When $d_1$ approaches $d_2$ we can see the rise of the $I_1$ amplitude due to the geometric resonance. Because of that, the first harmonic is dominant for equal ferromagnetic layers and the spin singlet and spin triplet with zero spin projection correlations practically generate the supercurrent~\cite{pajovic_josephson_2006}. In ferromagnetic bilayers only even harmonics (the second is the largest) can be generated by long-range spin-triplet correlations with $\pm 1$ spin projections~\cite{trifunovic_long-range_2011}.

The characteristic oscillations of $I_1(d_1), I_2(d_1)$, and $I_c(d_1)$ are due to $0-\pi$ transitions with the period practically equal to the ferromagnetic coherence length $\xi_F=20 k_F^{-1}$. Note that in the clean limit the critical current $I_c$ is minimum but not zero at the $0-\pi$ transition~\cite{Ryazanov_2001,Kontos_2001,radovic_josephson_2003}.

\subsection{Finite interfacial transparencies}
\begin{figure}[t!]
	\begin{center}
		\includegraphics[scale=0.97]{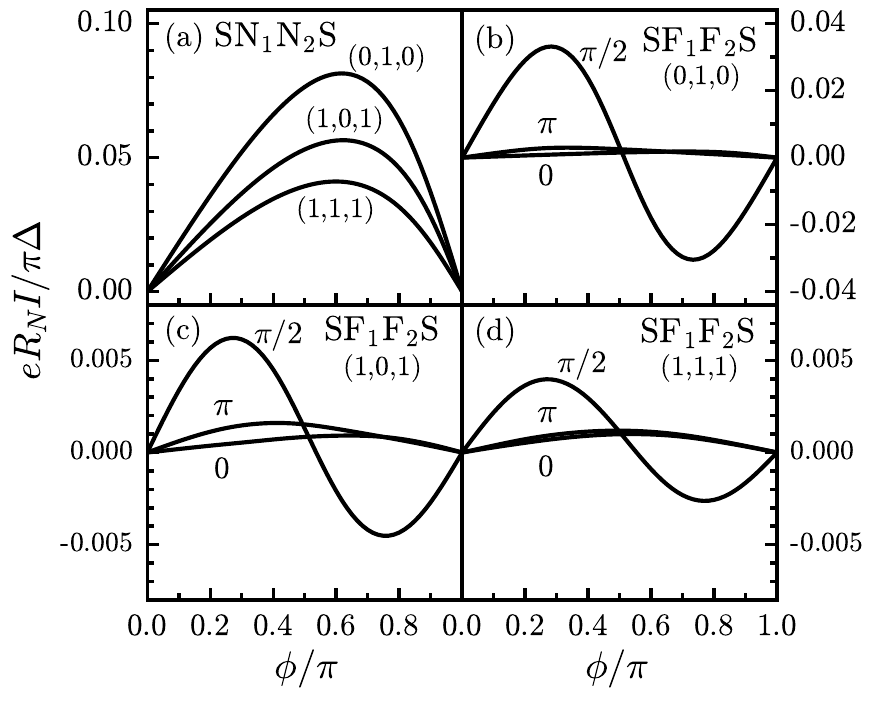}
	\end{center}
	\caption{
		The Josephson current in (a) $\S \N_1 \N_2 \S$ and (b)--(d)
		$\S \F_1 \F_2 \S$ junctions with layer thicknesses $d_1 = 10 k_F^{-1}$,
		$d_2 = 990 k_F^{-1}$, and different barrier strengths $(Z_1, Z_2, Z_3)$
		at the interfaces. Relative angles between the magnetizations
		$\alpha_r = 0$, $\pi/2$, $\pi$ in $\S \F_1 \F_2 \S$
		junctions are indicated in the plots.
	}
	\label{fig:allZ2}
\end{figure}
\begin{figure}[t!]
	\begin{center}
		\includegraphics[scale=1]{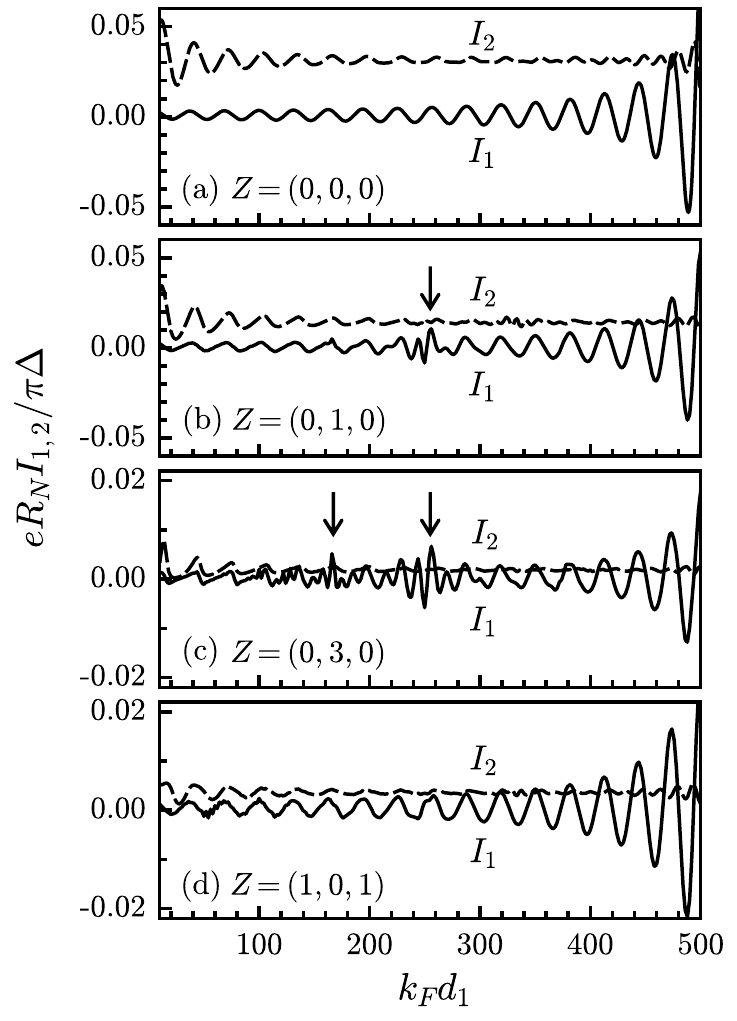}
	\end{center}
	\caption{
		The first harmonic amplitude (solid line) and the second harmonic amplitude (dashed line)
		of the Josephson current-phase relation in $\S\F_1\F_2\S$
		junctions with orthogonal magnetizations $\alpha_r=\pi/2$ as a
		function of the $\F_1$ layer thickness $d_1$, for total thickness $d_1+d_2 = 1000k_F^{-1}$,
		and for different barrier strengths at the interfaces $Z=(Z_1,Z_2,Z_3)$:
		(a) $Z=(0,0,0)$, (b) $Z=(0,1,0)$, (c) $Z=(0,3,0)$, and
		(d) $Z=(1,0,1)$. Additional geometric resonances are pointed to by arrows: (b) and (c).}
	\label{fig:I_HarmonicsZ}
\end{figure}
\begin{figure}[t!]
	\begin{center}
		\includegraphics[scale=1]{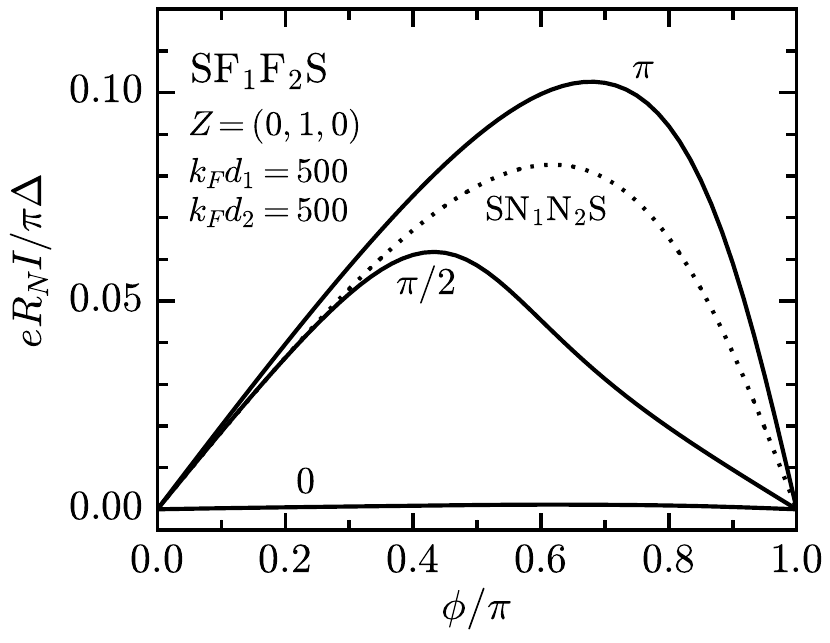}
	\end{center}
	\caption{
		The Josephson current in $\S \F_1 \F_2 \S$ junctions with
		equal layer thicknesses $d_1 = d_2 = 500 k_F^{-1}$
		and interfacial barrier strengths $Z_2=1$, $Z_1 = Z_3 = 0$, shown
		for different relative angles between magnetizations $\alpha_r =0$,
		$\pi/2$, $\pi$. The Josephson current in $\S \N_1 \N_2 \S$
		junction with the same layer thicknesses and barrier strengths
		is shown for comparison (dotted line).
	}
	\label{fig:ResonanceZ010}
\end{figure}

The role of finite interfacial transparencies is illustrated in Figs.~\ref{fig:allZ2}-\ref{fig:ResonanceZ010}. For comparison, the current-phase relation for a clean $\S\N_1\N_2\S$ $(h=0)$ junction with  $k_Fd_1=10$, $k_Fd_2=990$ is shown for various interfacial barrier strengths, see Fig.~\ref{fig:allZ2}(a). With decreasing transparency the supercurrent is suppressed in comparison to the fully transparent case [see  the dotted curve in Fig.~\ref{fig:IvsPhi}(a)]. In this case the first harmonic is dominant. The supercurrent of $\S\F_1\F_2\S$ junctions with $\alpha_r=0,\pi/2, \pi$ and different interfacial transparencies is shown in Figs.~\ref{fig:allZ2}(b)-(d). Note that for collinear magnetizations, $\alpha_r=0,\pi$, the current is short ranged and for orthogonal magnetizations, $\alpha_r=\pi/2$, the dominant second harmonic is due to the long-range spin-triplet correlations. It can be seen that a lower transparency of the interface between ferromagnets is less destructive than lower transparencies of the interfaces between superconductors and neighboring ferromagnetic layers. The depairing effect of normal reflection at the SF interfaces is stronger due to the direct suppression of the Andreev process.

The influence of finite interfacial transparencies on the first and the second harmonics is quite different. A first harmonic is generated by the phase-coherent transport of one Cooper pair, while the second harmonic is determined by the phase-coherent transport of two Cooper pairs. In Fig.~\ref{fig:I_HarmonicsZ} the first harmonic amplitude (solid curve) and the second harmonic amplitude (dashed curve) are shown as functions of $d_1$ for $k_F(d_1 + d_2) = 1000$, $\alpha_r=\pi/2$, and different transparencies of the interfaces, $Z=(Z_1,Z_2,Z_3)$. It can be seen that both $I_1$ and $I_2$ amplitudes are suppressed by decreasing the transparency of the interfaces, the first harmonic amplitude being much less affected. 

New geometric resonances and amplifications of $I_1$ emerge for a finite transparency of the interface between ferromagnetic layers [see Figs.~\ref{fig:I_HarmonicsZ}(b) and \ref{fig:I_HarmonicsZ}(c)]. Besides the resonant amplification of $I_1$ for $d_1=d_2$, we find resonant amplifications at $d_1=d_2/3, d_2/5, \dots$. This effect is related to the multiple reflections that lead to the emergence of electron and hole quasiclassical trajectories with a canceled phase accumulation.

The current-phase relations for equal ferromagnetic layers and finite interfacial transparency between them are shown in Fig.~\ref{fig:ResonanceZ010}. The critical currents are approximately two times smaller than in the fully transparent case [see Fig.~\ref{fig:IvsPhi}(a)]. We can see a peculiar amplification of the Josephson current for antiparallel magnetizations, $\alpha_r=\pi$,  in comparison with nonmagnetic layers. This effect was previously reported for SFIFS Josephson junctions with antiparallel orientations of magnetizations~\cite{Krivoruchko2001}, and for junctions between superconductors with ferromagnetic exchange fields~\cite{Bergeret_enhanced,Chtchelkatchev2002}. 

\section{Conclusions} \label{sec:Conclusion}
We have studied the Josephson effect in clean planar $\S\F_1\F_1\S$ junctions with arbitrary transparencies of the interfaces between the layers. By solving the scattering problem for the Bogoliubov-de Gennes equation, we have calculated numerically the current-phase relation, the critical current, and first and second harmonic amplitudes. For relatively a weak exchange field, $h/E_F = 0.1$, mutually orthogonal magnetizations, $\alpha_r=\pi/2$, and very unequal thicknesses of the ferromagnetic layers, $d_1\ll d_2$,  a well-pronounced second harmonic is obtained as a signature of the long-range spin-triplet correlations. On the other hand, for equally thick ferromagnetic layers, $d_1=d_2$, the  spin-singlet contribution to the first harmonic is enhanced due to the geometric resonance, and dominates even for thick layer junctions (strong ferromagnetic influence) with orthogonal magnetizations. 

Both resonant and spin-triplet effects qualitatively persist in the presence of impurities or moderate disorder (see, for example, the quasiclassical analysis in Refs.~\cite{trifunovic_josephson_2011,knezevic_signature_2012,richard_superharmonic_2013}). In experiments the resonances can be recognized as more sinusoidal $I(\phi)$, while the long-range spin-triplet correlations in the Josephson junctions with ferromagnetic bilayers lead to the more anharmonic current-phase relation due to the dominant second harmonic.

Both the first and the second harmonic amplitude show characteristic oscillations with varying thicknesses of the ferromagnetic layers. The critical current oscillates in the same manner. This is due to the $0-\pi$ transitions and the period of oscillations is the ferromagnetic coherence length $\xi_F$.

For a finite transparency of interfaces the supercurrent is suppressed, with higher harmonics being more affected. A low transparency of the $\F_1/\F_2$ interface, where the long-range spin-triplet correlations are generated, has a nontrivial impact on the interference phenomena and consequently to the current-phase relation. For certain thicknesses of the ferromagnetic layers in addition to $d_1=d_2$ new geometric resonances occur at $d_1=d_2/3, d_2/5, \dots$, making the first harmonic dominant even in asymmetric junctions. 

\acknowledgments
The work was supported by Serbian Ministry of Education, Science and Technological Development, Project No. 171027. Z.R. also acknowledges the support of the Serbian Academy of Sciences and Arts, Grant No. F87. A.I.B. acknowledges support by the Ministry of Science and Higher Education of the Russian Federation within the framework of state funding for the creation and development of World-Class Research Center  “Digital Biodesign and Personalized Healthcare”, Grant No. 075-15-2022-304. D.N. acknowledges support from Deutsche Forschungsgemeinschaft  (German Research Foundation) via SFB 1432 (Project No. 425217212).

\end{document}